\documentclass[fleqn,usenatbib]{mnras}

\usepackage{newtxtext,newtxmath}

\usepackage[T1]{fontenc}

\DeclareRobustCommand{\VAN}[3]{#2}
\let\VANthebibliography\thebibliography
\def\thebibliography{\DeclareRobustCommand{\VAN}[3]{##3}\VANthebibliography}

\usepackage{graphicx}	
\usepackage{amsmath}	
\usepackage{hyperref}
\usepackage{xcolor}

\newcommand{\sh}{SH}
\newcommand{\hl}{HL}

\usepackage[normalem]{ulem}

\title[Bias on Tensor-to-Scalar Ratio Inference]{Bias on Tensor-to-Scalar Ratio Inference With Estimated Covariance Matrices}

\author[D. Beck et al.]{
Dominic Beck$^{1,2}$\thanks{E-mail: dobeck@stanford.edu}, 
Ari Cukierman$^{1,2}$, 
W. L. Kimmy Wu$^{2}$
\\
$^{1}$Department of Physics, Stanford University, Stanford, California 94305, USA\\
$^{2}$Kavli Institute for Particle Astrophysics and Cosmology, SLAC National Accelerator Laboratory, 2575 Sand Hill Rd, Menlo Park, California 94025, USA\\
}

\date{Accepted XXX. Received YYY; in original form ZZZ}

\pubyear{2021}

\begin{document}
\label{firstpage}
\pagerange{\pageref{firstpage}--\pageref{lastpage}}
\maketitle

\begin{abstract}
We investigate simulation-based bandpower covariance matrices commonly used in cosmological parameter inferences such as the estimation of the tensor-to-scalar ratio~$r$. We find that upper limits on~$r$ can be biased low by tens of percent. The underestimation of the upper limit is most severe when the number of simulation realizations is similar to the number of observables. Convergence of the covariance-matrix estimation can require a number of simulations an order of magnitude larger than the number of observables, which could mean~$\mathcal{O}(10\ 000)$ simulations.
This is found to be caused by an additional scatter in the posterior probability of~$r$ due to Monte Carlo noise in the estimated bandpower covariance matrix, in particular, by spurious non-zero off-diagonal elements. We show that matrix conditioning can be a viable mitigation strategy in the case that legitimate covariance assumptions can be made.
\end{abstract}




\section{Introduction}

Measuring primordial gravitational waves (PGWs) predicted by models of cosmic inflation is not only a major experimental challenge but also a demanding problem in statistical inference in the realm of cosmological parameter estimation. Current and future constraints on the tensor-to-scalar ratio~$r$, which quantifies the power of PGWs, are driven by measurements of so-called $B$~modes in the polarization of the cosmic microwave background~(CMB) \citep{Seljak1997,Kamionkowski1997}.
The predicted primordial $B$-mode signal is faint compared with current instrumental sensitivities, $B$-modes from weak gravitational lensing and astrophysical foregrounds \citep{Kamionkowski2016}. For this reason, setting constraints on the parameter~$r$ can be different from the  high-signal-to-noise measurements of the six $\Lambda$CDM parameters from CMB temperature and $E$-mode polarization. Due to the profound implications of a potential detection of non-zero~$r$, careful data analysis is warranted at every level, including that of parameter inference. \\

At the time of writing, the strongest constraint on~$r$ is set by the BICEP/\textit{Keck} Collaboration through measurements of the so-called \emph{recombination peak} in the $B$-mode spectrum at multipoles $\ell \gtrsim 30$. With data gathered up to and including the 2018 observing season, BICEP/\textit{Keck} reported an upper limit of $r_{0.05}<0.036$ at 95\% confidence level (95\% C.L.) with a fixed $\Lambda$CDM cosmological model from \textit{Planck} 2018 \citep{BK18,Planck18}, where the subscript~$0.05$ refers to the \emph{pivot scale} defined in Sec.~\ref{sec:parest}. We refer to this result as ``BK18.'' The uncertainty on~$r$, which we denote~$\sigma_r$, is measured from a set of simulations by calculating the standard deviation of the maximum-likelihood estimates of~$r$. For the baseline BK18 data set, this results in $\sigma_r=0.009$. In combination with baryon acoustic oscillations (BAO) and \textit{Planck} 2018 temperature and $E$-mode polarization data \citep{bao,Planck18}, the constraint tightens to $r_{0.05}<0.035$ (95\% C.L.) after marginalizing over the six $\Lambda$CDM parameters including~$n_s$, the spectral index of the scalar power spectrum, and~$\tau$, the optical depth to reionization \citep{BK18}. \\

At multipoles smaller than those accessible by BICEP/\textit{Keck}, temperature and polarization data from the \textit{Planck} satellite mission can be used to constrain~$r$ \citep{Tristram20,Tristram21}. The \textit{Planck} large-scale polarization data are derived with the NPIPE processing pipeline \citep[][PR4]{Planck20}, which produces calibrated frequency maps with leading sensitivity on the full sky. 
In \cite{Tristram20}, the NPIPE polarization maps were combined with high-$\ell$ CMB temperature data, and an upper limit of $r_{0.05} < 0.056$ (95\% C.L.) was reported. In \cite{Tristram21} the authors utilize the updated BK18 likelihood, reporting an upper limit of $r_{0.05} < 0.032$ (95\% C.L.) after making modifications to their analysis \citep{Tristram21}.
\\

All mentioned analyses inferring an upper limit on $r$ are relying on the construction of a covariance matrix from a set of simulations of a finite number. In this paper, we address the problem of using a limited number of simulations to estimate a covariance matrix to account for statistical and systematic uncertainties in the data.
From previous studies of cosmological data, it is known that an ill-formed covariance matrix can cause parameter estimates to be biased and uncertainties to be underestimated \citep{Hartlap06,HL08,Taylor2013,Dodelson2013,Balkenhol21,Percival2021}. 
A common remedy is the conditioning of the covariance matrix, which can involve masking most off-diagonal matrix elements or applying some semi-analytical corrections (see \cite{Balkenhol21} for a review). 
These conditioning strategies are followed to varying extents in $r$~inferences using data from ground-based experiments \citep{pbr,sptr,spider21,BK18}. 
In \cite{Hartlap06} a de-biasing of the estimated inverse covariance matrix is proposed, which effectively widens the Gaussian likelihood but does not account for the randomness of the estimated covariance matrix. 
Assuming the estimated covariance matrix is Wishart distributed, one can marginalize analytically over the randomness of the covariance-matrix estimate, resulting in a replacement of the Gaussian likelihood with a modified multivariate $t$~distribution \citep{Sellentin15}. We will address the utility and limits of this likelihood correction in the context of $r$~parameter inference.\\

Furthermore, we show in Sec.~\ref{sec:peakscatter} that using a finite number of simulations to generate the covariance matrix can lead to an additional scatter in the maximum \textit{a posteriori} (MAP) estimate of~$r$.
We then present in Sec.~\ref{sec:uncertain} that the known
underestimation of parameter uncertainties can be mitigated by
marginalizing over the randomness of the covariance matrix or employing a tuned prior \citep{Percival2021}.
However, even with a properly accounted uncertainty, the noise in the covariance matrix still leads to misestimations of the upper limit.
We demonstrate in Sec.~\ref{sec:clscatter} that the additional scatter in the MAP estimate can cause a significant misestimate and likely an \textit{underestimate} of an upper limit on~$r$.
We note that the PR4-based upper limits reported in \cite{Tristram20} and \cite{Tristram21} can be affected by such biases.

\section{Toy-Model Simulations}\label{sec:toymodel}

We run numerical experiments to assess the impact of using a finite number of simulation realizations to estimate a covariance matrix. Each simulation includes the following components: 
\begin{itemize}
    \item A Gaussian realization of $\Lambda$CDM CMB consistent with lensed power spectra following the best-fit model of \textit{Planck} 2018 \citep{Planck18},
    \item An additional tensor component with $r=0.01$,
    \item A Gaussian realization of noise with a power spectrum given by
\end{itemize} 
\begin{equation}
    N_\ell=\left(1+\left(\frac{\ell}{\ell_{\rm knee}}\right)^\alpha\right)\sigma^2 b_\ell^{-2},
\end{equation}
with parameters $\sigma=60\ \mu \mathrm{K} \textrm{-arcmin}$, $\ell_{\rm knee}=10$, $\alpha=-3$ and a Gaussian beam function $b_\ell$ with $30\ \textrm{arcmin}$ FWHM. The arguments in this paper are robust to the particular choices for these parameters and focus on effects caused by the \textit{relative} scatter of the covariance-matrix estimate. 
Our simulations are idealized in that they omit non-Gaussian effects from, e.g., masking, realistic noise, instrumental systematics, astrophysical foregrounds, gravitational lensing, etc. Hence, the underlying sky maps in these simulations are Gaussian random fields. Hence we can assume that the true underlying covariance of the bandpowers computed on the full sky is zero between different multipoles for $\ell \lesssim 1000$, i.e., for scales larger than the pixel size ($\sim 7\ \textrm{arcmin}$). 
For each full-sky map, we compute $EE$, $EB$ and $BB$ power spectra and employ the multipole binning of \cite{Tristram20}, amounting to 45 bins per spectrum in the multipole range~$\ell\in[2,150]$. When using all three spectra, then, the total number of observables is 135.

\section{Parameter Estimation}\label{sec:parest}

To infer a tensor-to-scalar ratio~$r$ from CMB $E$ and $B$ modes in our toy model, we assume a one-parameter model, defining $r$ as the ratio of the tensor and scalar power spectra at a pivot scale $k_*=0.05\ {\rm Mpc}^{-1}$. The likelihood is based on \cite{HL08}, which accounts for the non-Gaussian distribution of bandpowers by constructing a Gaussian likelihood approximation in the derived variables
\begin{equation}
    X_b={\rm vecp} \left( \sqrt{C_b^{\rm fid.}} g\left( \sqrt{C_b}^{-1} \hat{C}_b \sqrt{C_b}^{-1} \right) \sqrt{C_b^{\rm fid.}} \right), \label{eq:defineX}
\end{equation}
where $C_b^{\rm fid.}$~are fiducial binned theoretical spectra following the same model used in the simulations, $\hat{C}_b$~are the data bandpowers, and $C_b$~are the model bandpowers (including a potential noise bias or offset). See Appendix~\ref{sec:appendix} for the definitions of ${\rm vecp}$ and~$g$. This likelihood, which we label by ``\hl,'' is given by 
\begin{equation}
    \mathcal{L}_\textrm{HL} = \frac{1}{\sqrt{2\pi|\mathbf{M}|}}\exp\left( - \frac{1}{2} \mathbf{X}^T \mathbf{M}^{-1} \mathbf{X} \right),
    \label{eq:hl}
\end{equation}
where $\mathbf{M}$~is the bandpower covariance matrix.
In the next section, we will focus on the construction of~$\mathbf{M}$. \\

At several points, we will compare the \hl\ likelihood to the correction obtained by marginalizing over the distribution of the sample covariance matrix (the covariance matrix estimated from simulations). Following \cite{Sellentin15}, the corrected likelihood replacing Eq.~\ref{eq:hl} that accounts for the statistical distribution of a simulated covariance matrix $\mathbf{M}$ is given by
\begin{equation}
    \mathcal{L}_\textrm{SH} = \frac{c}{\sqrt{2\pi|\mathbf{M}|}} \left( 1+ \frac{\mathbf{X}^T \mathbf{M}^{-1} \mathbf{X}}{N_{\rm sims}-1} \right)^{-\frac{N_{\rm sims}}{2}},
    \label{eq:sh}
\end{equation}
which we label as ``SH'' likelihood. See Eq.~\ref{eq:gammafct} for the expression of the normalization constant $c$. We evaluate the single-parameter posteriors with uniform priors on~$r$.

\begin{figure*}
\centering
\includegraphics[width=\linewidth]{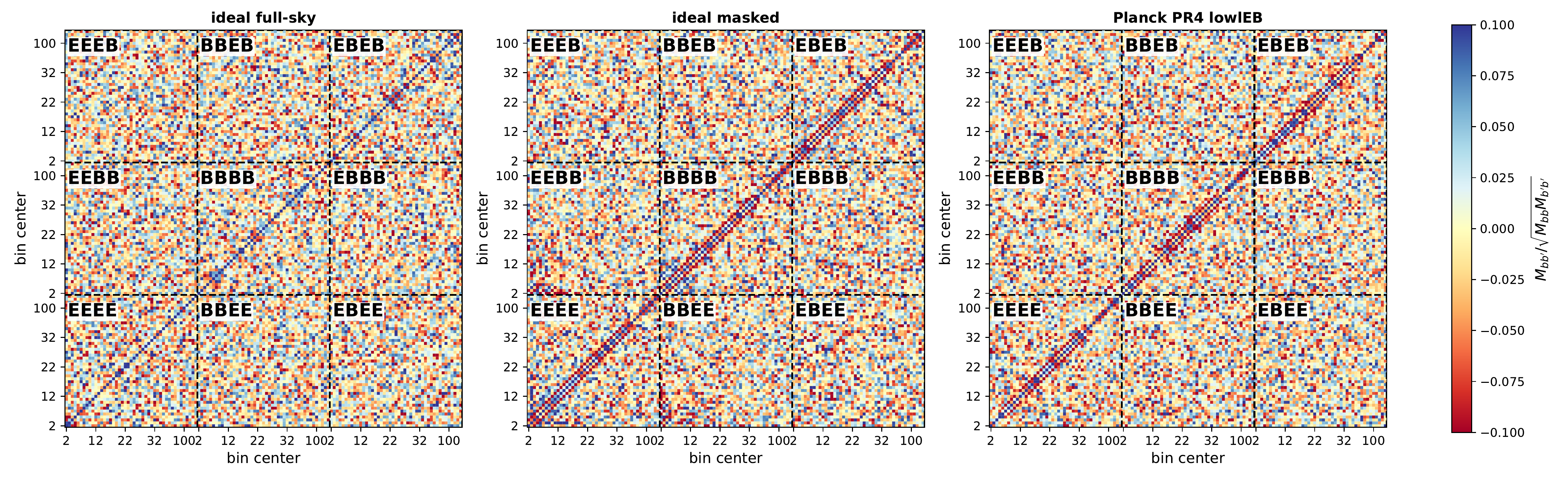}
\caption{Left: Correlation matrix (covariance matrix normalized to the diagonal) of the toy-model simulations described in Sec.~\ref{sec:toymodel}, using $400$ simulations as in \protect\cite{Tristram20,Tristram21}. 
Center: Same as the correlation matrix on the left, but the maps have been masked with an $f_{\rm sky}=50\%$ Galaxy mask before computing purified power spectra with \texttt{NaMaster}.
Right: Correlation matrix of the Planck ``\textit{lowlEB}'' likelihood for $EE$, $EB$ and $BB$ bandpowers and employing the same binning as described in \protect\cite{Tristram21}.}  \label{fig:toysimcov}
\end{figure*}

\section{Covariance Matrix Estimation and Conditioning}\label{sec:covmat}

An unbiased estimator of the covariance matrix between bandpowers~$C_b$ is
\begin{equation}
    \hat{\mathbf{M}}_{bb'}=\frac{1}{N_{\rm sims}-1} \sum_{i=1}^{N_{\rm sims}} \left( \hat{C}^{(i)}_b - \left\langle \hat{C}^{(i)}_b \right\rangle \right) \left( \hat{C}^{(i)}_{b'} - \left\langle \hat{C}^{(i)}_{b'} \right\rangle \right),
    \label{eq:covmatest}
\end{equation}
where $\hat{C}^{(i)}_b$~is the bandpower in bin $b$ for the $i$th~independent simulation. The total number of independent simulations is~$N_{\rm sims}$. Using the simulations described in Sec.~\ref{sec:toymodel}, we can construct mock covariance matrices for any given number of simulations. On the left-hand side of Fig.~\ref{fig:toysimcov}, we show a correlation matrix estimated from $EE$, $EB$ and $BB$ spectra computed from 400 simulations of CMB and noise as described in Sec.~\ref{sec:toymodel}. \\

Invoking the central limit theorem that allows us to assume that the bandpowers $\hat{C}^{(i)}_b$ are approximately Gaussian distributed, we can assume that the estimated covariance matrix~$\hat{\mathbf{M}}$ follows a Wishart distribution given the true covariance matrix~$\mathbf{M}$ \citep{anderson2003}
\begin{equation}
    P(\hat{\mathbf{M}}|\mathbf{M},N_{\rm sims}) \propto |\mathbf{M}|^{\frac{N_{\rm sims}-p-2}{2}} e^{-\frac{1}{2}\left(N_{\rm sims}-1\right) {\rm Tr} \left( \mathbf{M}^{-1}\hat{\mathbf{M}} \right) },
    \label{eq:wishart}
\end{equation}
where $p$~is the number of bandpowers, i.e., the length of the vector~$\mathbf{X}$. Samples of this distribution converge to the true covariance matrix for an infinite number of simulations~$N_{\rm sims}$. 
For a finite number of simulations, each matrix element contains a random statistical component that we will call Monte Carlo noise (MC noise).
The MC noise can be observed in the off-diagonal elements of the matrix presented on the left of Fig.~\ref{fig:toysimcov}. Since the underlying true covariance matrix is diagonal, any non-zero off-diagonal element is due to MC noise.\\

On the right-hand side of Fig.~\ref{fig:toysimcov}, we show a practical example of a binned correlation matrix estimated using more realistic simulations provided within the \textit{Planck} PR4 ``\textit{lowlEB}'' likelihood. This likelihood is based on foreground-cleaned CMB polarization maps produced using \textsc{COMMANDER} on \textit{Planck} PR4 frequency maps \citep{Tristram21}. The $EE$, $EB$ and $BB$ spectra are compared to the model theory through the \hl\ likelihood approximation, which fits--in the nature of the \hl\ approximation--deviations from a fiducial set of bandpowers to their $\Lambda$CDM expectation values given a covariance matrix estimated from Monte Carlo simulations of the fiducial model, noise and systematics. The likelihood is publicly available.\footnote{\texttt{https://github.com/planck-npipe/lollipop}} Figure~\ref{fig:toysimcov} shows the corresponding bandpower correlation matrix as it is used by the likelihood code. \\

We find that the distribution of off-diagonal elements in the \textit{Planck} ``\textit{lowlEB}'' covariance matrix (rightmost in Fig. \ref{fig:toysimcov}) is consistent with our toy model (leftmost in Fig. \ref{fig:toysimcov}) outside of the obvious structure on the diagonal and neighboring elements. Correlations near the diagonal are expected in a covariance matrix estimated from Monte Carlo simulations that include masking, realistic instrumental noise, \textit{Planck} systematic effects incorporated in the PR4 simulations \citep{Planck20}, component-separation uncertainties and foreground residuals. \\

\begin{figure}
\centering
 \includegraphics[width=\linewidth]{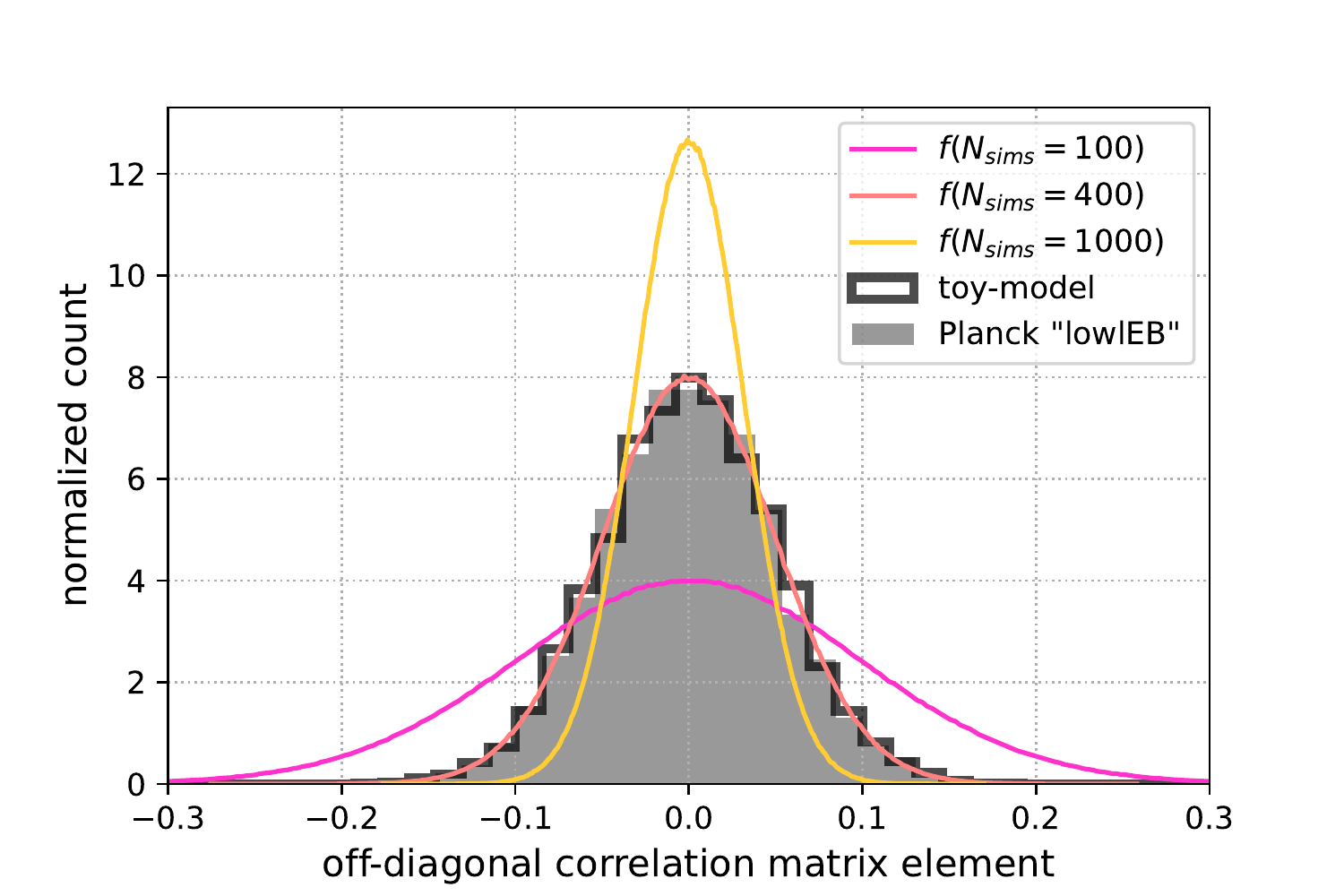}
\caption{Histogram of the off-diagonal elements of the correlation matrix, i.e., the covariance-matrix elements normalized to the diagonal, for the ideal simulations described in Sec.~\ref{sec:toymodel} for $N_{\rm sims}=400$ (Fig.~\ref{fig:toysimcov} left) compared to the off-diagonal covariance-matrix elements of the \textit{Planck} ``\textit{lowlEB}'' likelihood (Fig.~\ref{fig:toysimcov} right). The solid lines show the marginal distributions~$f(N_{\rm sims})$ of off-diagonal elements for Wishart-distributed covariance matrices. 
} \label{fig:toysimhist}
\end{figure}

In the center of Fig.~\ref{fig:toysimcov}, we show the correlation matrix for ideal simulations with a 50\%~Galaxy mask, where the power spectra are estimated with the pure pseudo-$C_\ell$ estimator {\tt NaMaster}\footnote{\texttt{https://github.com/LSSTDESC/NaMaster}} \citep{namaster}. Despite the difference in power-spectrum estimators, we can reproduce the superficial structure in the \textit{Planck} PR4 ``\textit{lowlEB}'' correlation matrix (rightmost in Fig. \ref{fig:toysimcov}). 
In Fig.~\ref{fig:toysimhist}, we show the expected marginal distribution of off-diagonal elements from sampled Wishart matrices for $N_{\rm sims}=100,\ 400\ {\rm and}\ 1000$. 
We histogram the distribution of the off-diagonal correlation-matrix elements for the toy model with $N_{\rm sims}=400$ (Fig.~\ref{fig:toysimcov} left) and for matrix elements that go beyond obvious correlation structure in the \textit{Planck} ``\textit{lowlEB}'' likelihood (Fig.~\ref{fig:toysimcov} right). This includes all elements beyond next-to-nearest-neighbor correlations for low multipoles, where bandpowers are unbinned, and beyond nearest-neighbor correlations for binned bandpowers. Throughout the paper we will refer to a covariance matrix for which we set these elements to zero as ``conditioned.'' The \sh\ likelihood is, by construction, incompatible with this kind of matrix conditioning, and we will, therefore, use conditioned matrices only in the \hl\ likelihood. \\

The distributions corresponding to the toy-model simulations as well as the \textit{Planck} PR4 case are consistent with the expected marginal distribution of sampled off-diagonal Wishart matrix elements for $N_{\rm sims}=400$. A Kolmogorov–Smirnov (KS) test establishes the consistency of these distributions with a $p$~value of~12\% for the PR4 matrix and a $p$~value of~44\% for the toy-model matrix. These comparisons suggest that the off-diagonal matrix elements in the \textit{Planck} ``\textit{lowlEB}'' covariance matrix are consistent with MC noise. 
We note that non-Gaussian structure may still be hidden below the MC noise or could be latent in this particular statistic \citep{Braspenning21}. 

\section{Maximum a Posteriori Estimates}\label{sec:peakscatter}

\begin{figure}
\centering
 \includegraphics[width=\linewidth]{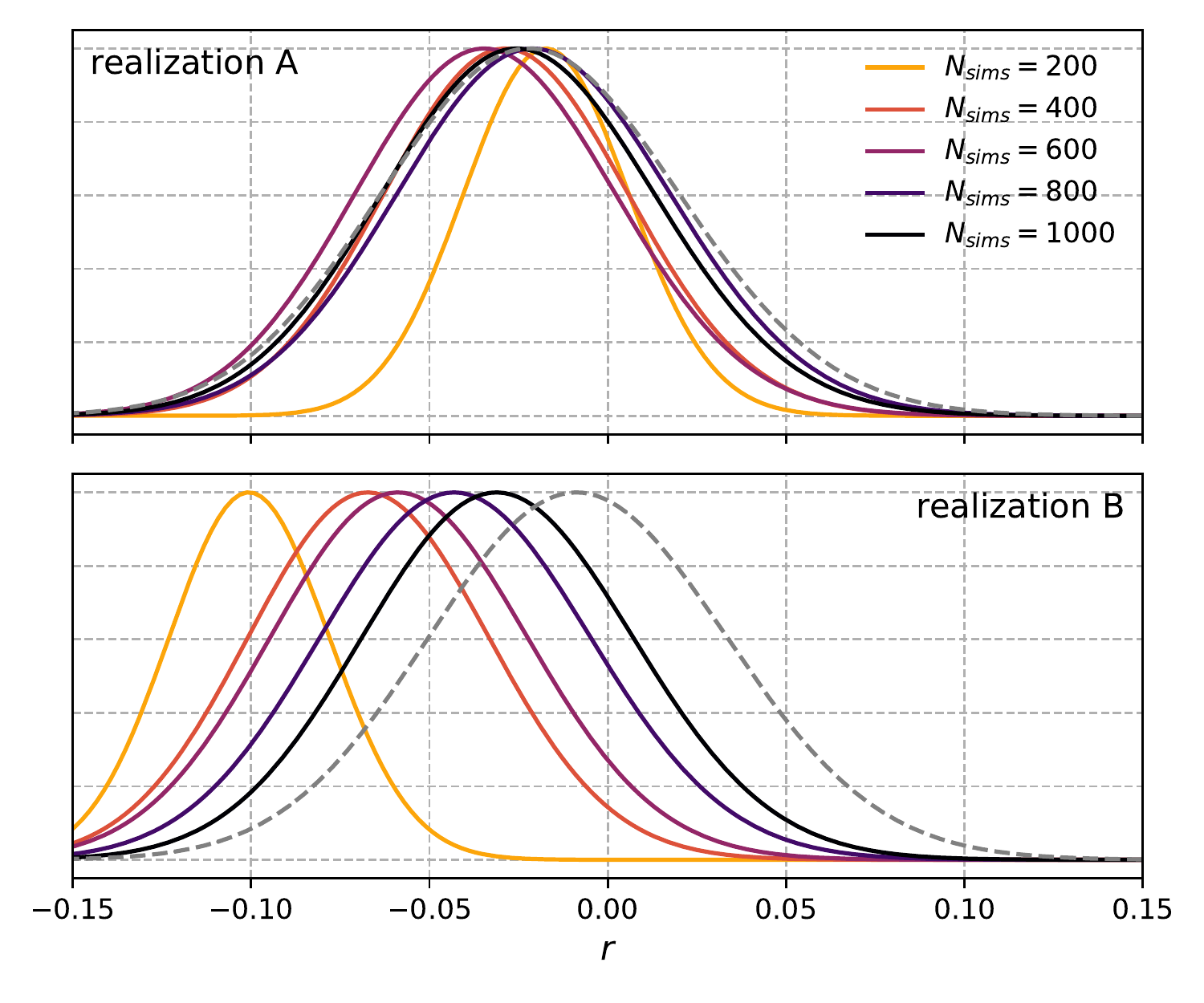}
\caption{Posterior densities of $r$ for two examples of simulated data realizations, which we refer to as realization A and B,  for a varying number of simulations $N_{\rm sims}$ used in the estimation of the covariance matrix. Here the \hl\ likelihood approximation is used with both $E$- and $B$-mode spectra. The gray dashed line is the posterior density using the ``true'' covariance matrix, estimated with $N_{\rm sims}=50\,000$.} \label{fig:toysimposteriors}
\end{figure}

We estimate the maximum \textit{a posteriori} (MAP) probability of~$r$ by maximizing the posterior densities for each of the 500 simulated ``data'' bandpowers. The simulated ``data'' vectors are non-overlapping with the set of simulations used for estimating the covariance matrix. We estimate covariance matrices (Eq.~\ref{eq:covmatest}) with a varying number of simulations and compare subsequent results obtained with the ``true'' covariance matrix, which we estimate with $N_{\rm sims}=50\,000$. In Fig.~\ref{fig:toysimposteriors}, we show two examples of posteriors from our 500 simulated data realizations, which are chosen to illustrate two extreme cases. In the first example (which we refer to as ``realization A'' in the figure), the posterior density converges quickly towards the true posterior with a relatively small number of simulations used for the covariance-matrix estimate. In the other case (referred to as ``realization B''), however, the posterior probability fluctuates for varying number of $N_{\rm sims}$, and we observe a significant misestimate of the posterior for $N_{\rm sims}<1000$ in both peak location and width.\\

We define a metric to measure the fidelity of the sample covariance matrix. For a given value of $N_\mathrm{sims}$ and for each of the 500 ``data'' simulations, we find the MAP estimate of~$r$, which we denote~$\hat{r}_{N_\mathrm{sims}}^{(i)}$, where $i$~labels the ``data'' realization. Then, for that specific realization, we calculate the deviation from the ``true'' MAP estimate, which we denote~$\hat{r}_{\mathrm{true}}^{(i)}$. Explicitly, the deviation is given by
\begin{equation}
    \Delta r^{(i)}_{\rm MAP}=\hat{r}_{N_\mathrm{sims}}^{(i)} - \hat{r}_{\mathrm{true}}^{(i)}.
    \label{eq:deltarmap}
\end{equation}
We normalize $\Delta r^{(i)}_{\rm MAP}$ by the statistical uncertainty on~$r$, which we denote $\sigma_r$ and which refers to the standard deviation of the MAP estimates from 500 ``data'' simulations with the ``true'' covariance matrix.\\

\begin{figure}
\centering
 \includegraphics[width=\linewidth]{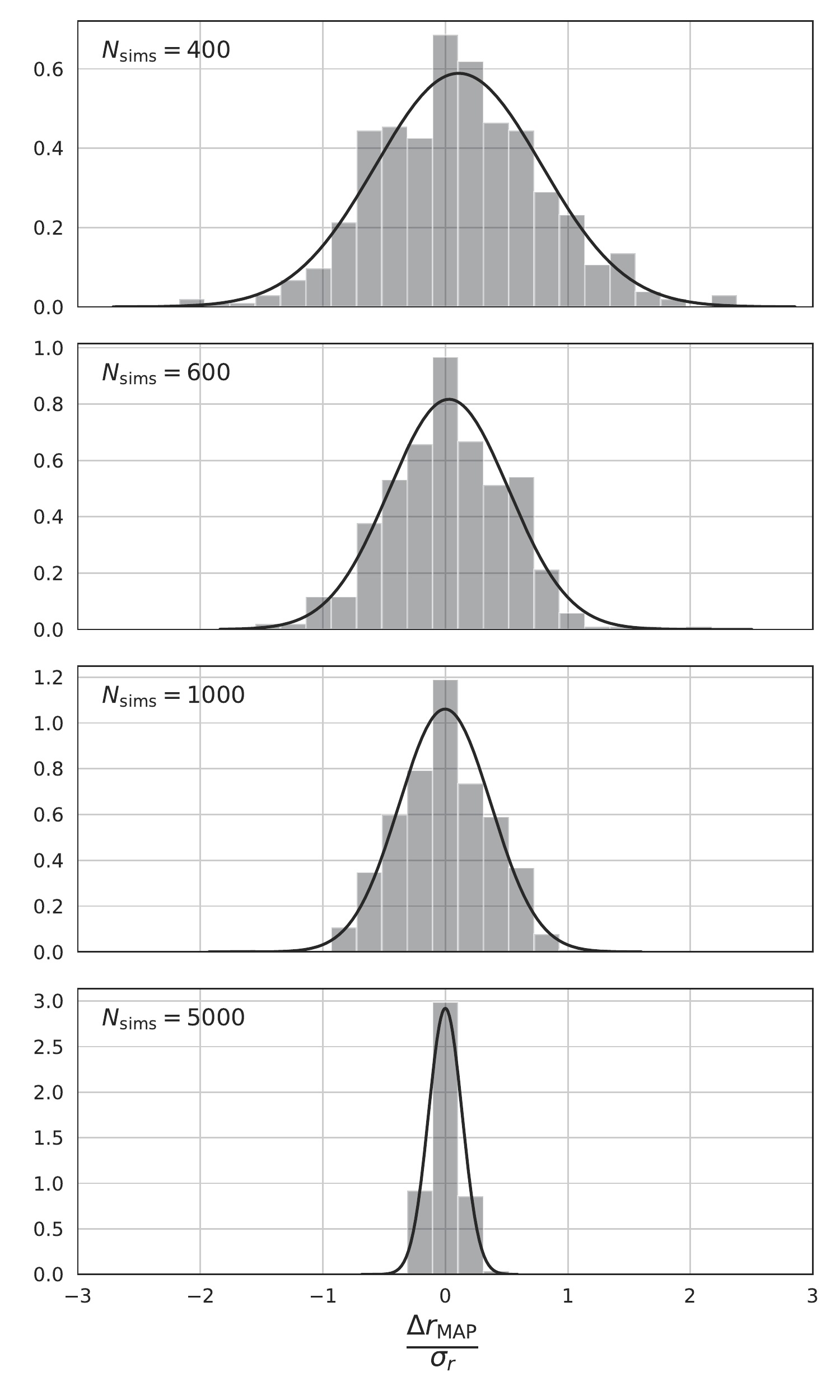}
\caption{The per-realization difference~$\Delta r^{(i)}_{\rm MAP}$ (Eq.~\ref{eq:deltarmap}) between MAP estimates of $r$ using a sample covariance matrix estimated with different values of $N_{\rm sims}$ and the fiducial ``true'' covariance using $N_{\rm sims}=50\,000$. The distribution of this difference for 500 realizations of input data vectors is shown in this figure, measured in units of the fiducial statistical uncertainty~$\sigma_r$. Although we do not expect these distributions to be exactly Gaussian, we show a Gaussian fit to each histogram in black solid lines.} \label{fig:mapsearch}
\end{figure}

In Fig.~\ref{fig:mapsearch}, we show the distribution of $\Delta r_{\rm MAP}$ for 500~simulations of the CMB. The MAP estimates are unbiased, but the scatter is larger when fewer simulations are used in the covariance-matrix estimate. The same effect has been described in \cite{Balkenhol21} as an additional scatter in parameter constraints that is unaccounted for. Note that this effect depends on the particular realization of the (simulated) data used to compute bandpowers and, therefore, is present in both \hl\ and \sh\ likelihood approximations, which peak at the same value of~$r$. \\

In the case of ideal toy-model simulations and combining $E$~and $B$~modes in the likelihood, 12\%~of posteriors for $N_{\rm sims}=400$ peak at~$1\sigma$ or further from the ``true'' peak. This is reduced to below~1\% for $N_{\rm sims}>1000$. \\

\begin{figure}
\centering
\includegraphics[width=\linewidth]{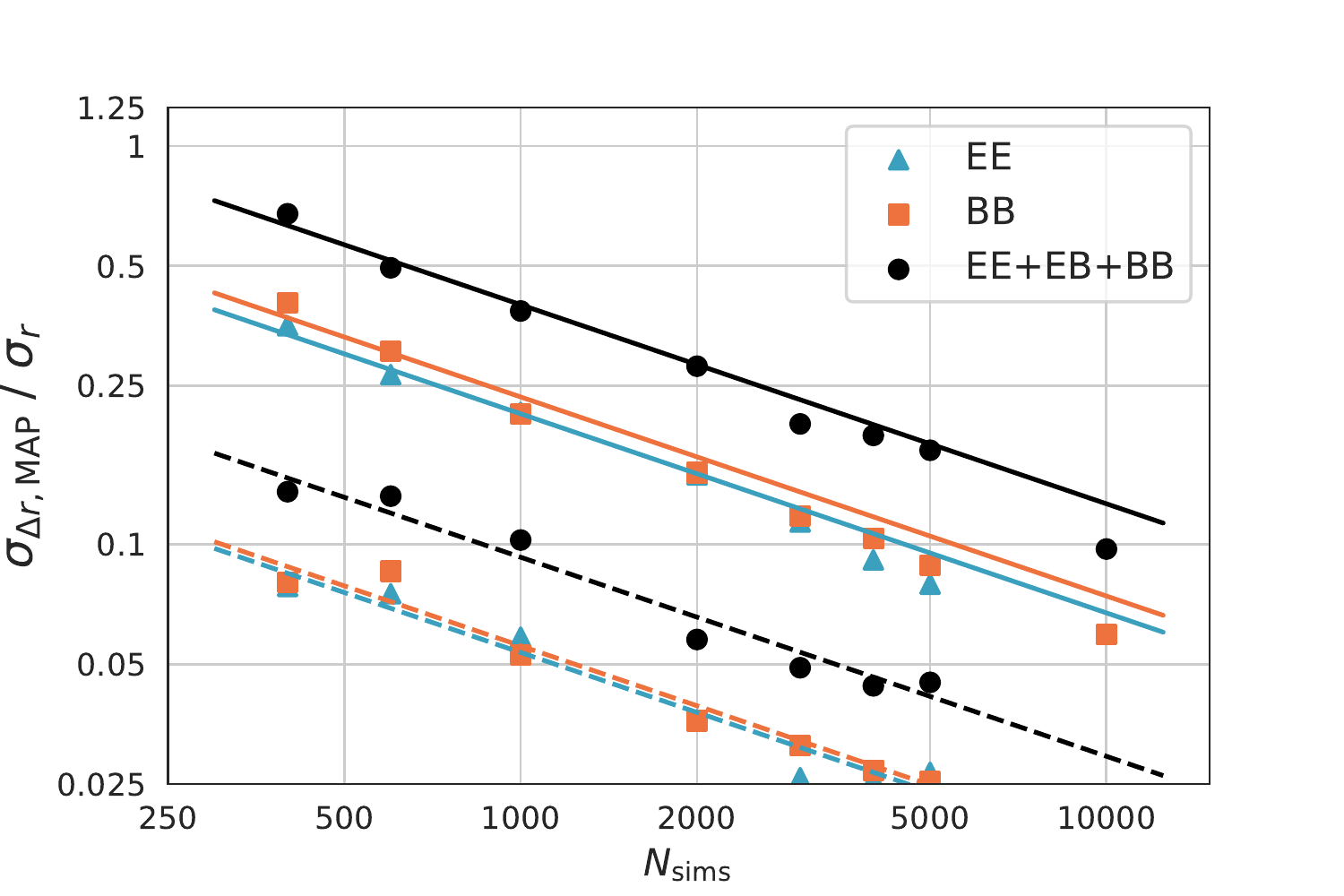}
\caption{The scatter of $\Delta r_{\rm MAP}$ shown in Fig.~\ref{fig:mapsearch} relative to the fiducial statistical uncertainty from our toy-model simulations as a function of $N_{\rm sims}$, the number of simulations used in the covariance-matrix estimator. The results inferred from the toy-model simulations are plotted as dots and overlayed with fits to $1/\sqrt{N_{\rm sims}}$ in continuous lines. The different colors indicate the type of spectra used in the likelihood, i.e., only $E$- or $B$-mode information or the combination of both. Solid lines indicate that the full covariance matrix was used in the likelihood. For the dashed lines, we condition the covariance matrix by setting off-diagonal elements to zero in order to isolate the effect from the diagonal. } \label{fig:mapsearch_vs_sims}
\end{figure}

In Fig.~\ref{fig:mapsearch_vs_sims}, we show as a function of~$N_{\rm sims}$ the standard deviation of~$\Delta r_{\rm MAP}$ relative to the fiducial uncertainty~$\sigma_r$. We find that this scatter is roughly proportional to $1/\sqrt{N_{\rm sims}}$. Reducing this scatter to below~10\% of the total fiducial statistical uncertainty requires $N_{\rm sims}\approx 10\,000$. This effect can be significantly mitigated by setting off-diagonal elements of the covariance matrix to zero (``conditioning''), which should, however, take place under consideration of possible covariance-inducing effects in the real data.

\section{Uncertainty Estimation}\label{sec:uncertain}

In this section, we address the issue of misestimating parameter uncertainties and credible intervals due to MC~noise in the covariance matrix. For this purpose, we compute a symmetric 68\% credible interval from the posterior of each simulated realization. We denote half the size of the 68\% credible interval as~$\sigma_\mathcal{L}$. We compare this estimate of the $1\sigma$~uncertainty with~$\sigma_r$, which is defined in the previous section and which is obtained by taking the standard deviation of MAP estimates using the fiducial ``true'' covariance matrix.\\

\begin{figure}
\centering
 \includegraphics[width=\linewidth]{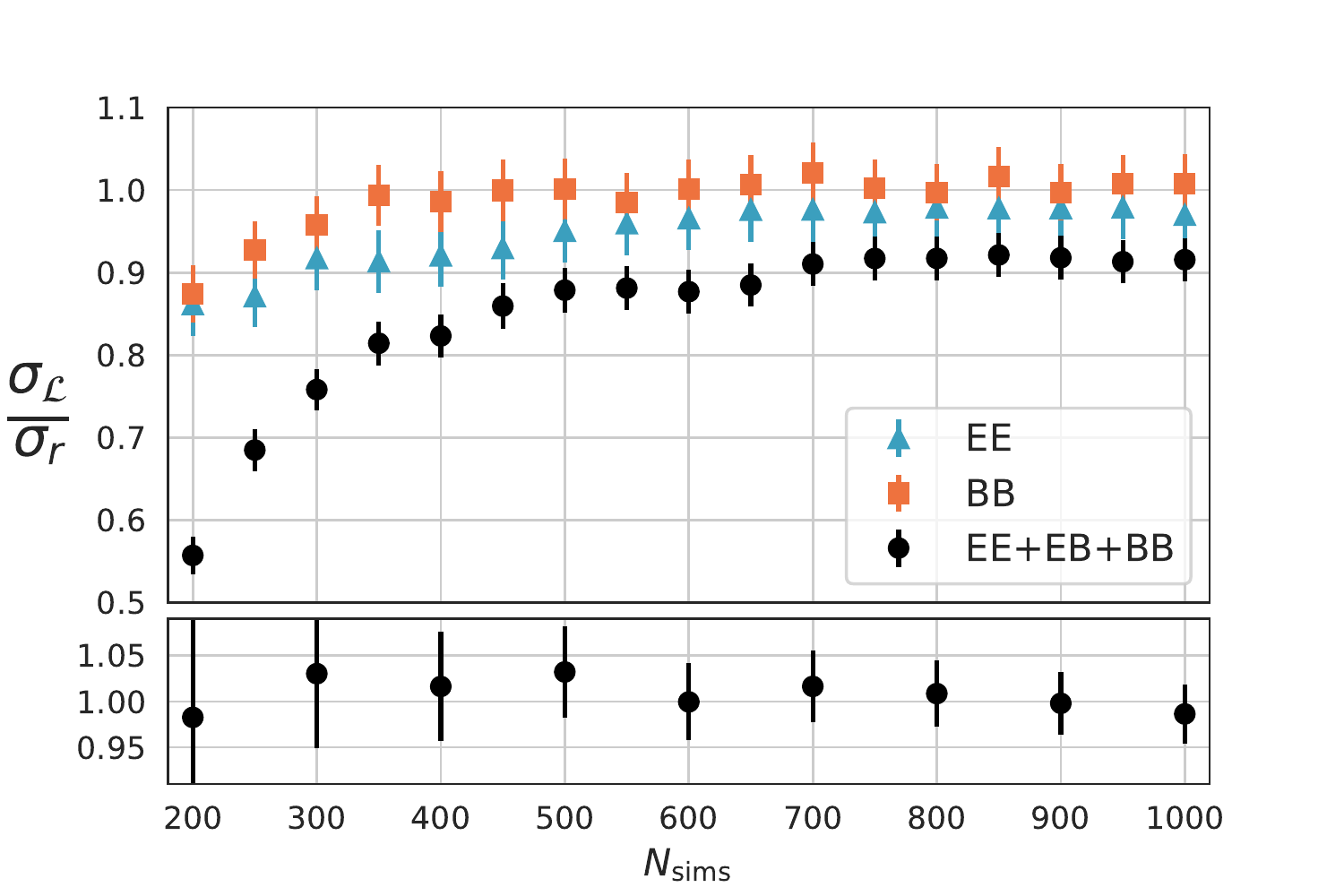}
\caption{
The ratio of $1\sigma$~uncertainties calculated in two ways: from the posterior width using $N_{\rm sims}$~simulations and from the scatter in MAP best-fit values using the true covariance matrix. This ratio should be unity, if the covariance-matrix estimate has converged.
The upper panel shows the ratio in the case of the \hl\ likelihood for either $E$- or $B$-mode-only bandpowers or their combination. The lower panel shows the ratio using the \sh\ likelihood correction. The error bars represent the standard deviation from the ensemble of 500~simulated data vectors.} \label{fig:uncertainty_vs_sims}
\end{figure}

In Fig.~\ref{fig:uncertainty_vs_sims}, we show the ratio of the two estimates and find an \textit{underestimation} of the \hl-likelihood credible intervals by~18\% for $N_{\rm sims}=400$. This effect is smaller in the individual $E$-mode-only or $B$-mode-only likelihoods, which include only a third as many observables as the combined likelihood. As shown in the lower panel of Fig.~\ref{fig:uncertainty_vs_sims}, this underestimate is effectively mitigated by the use of the \sh\ likelihood correction. \\
\\

A test to validate the faithfulness of the covariance estimate in a Bayesian framework is to compare the frequentist coverage probability, the proportion of simulation realizations that contain the true value of~$r$ in their credible intervals, to the nominal posterior quantile \citep{Cook2006}, e.g., the 68\% or 95\% credible intervals. As we show in Tab.~\ref{tab:credintervals}, the coverage probability using the \hl\ likelihood with a covariance matrix estimated from $N_{\rm sims}=400$ for our ``data'' simulation is~47\%~(84\%) for a nominal coverage probability of the credible interval of~68\%~(95\%). The \hl-based coverage probabilities are \emph{fractionally} smaller by 30\% (12\%), i.e., the fiducial true value of~$r$ lies within the confidence interval constructed for each realization 30\% (12\%) less often than the expected nominal posterior quantile. The coverage probabilities are matched to within~5\% for $N_{\rm sims}=2500$ ($N_{\rm sims}=700$). The \sh\ likelihood misestimates the confidence limit for $N_{\rm sims}=400$ by~15\%~(3\%). The remaining mismatch shows that, while the \sh\ likelihood significantly reduces the error in the estimation of the uncertainty as it was designed for (i.e., the posterior might have the correct width), it does not account for the extra scatter in the posterior peak position, as suggested in Sec.~\ref{sec:peakscatter}. 
In Tab.~\ref{tab:credintervals}, we also include the coverage probabilities from a posterior constructed with priors derived to match the frequentist confidence intervals~\citep[][PFSH]{Percival2021}. We note that the additional scatter in the MAP is accounted for in this case.
While the mismatch between frequentist coverage probabilites and Bayesian credible intervals is alleviated, this approach leads to overestimated limits on~$r$, as discussed in the next section.
In the case of the \hl\ likelihood approximation, we find that using more simulations in the covariance estimator or appropriate conditioning of the covariance matrix can improve the probability coverage matching.

\begin{table}
\caption{Covarage probabilities estimated from $500$~simulations corresponding to nominal 68\% (95\% in parentheses) credible intervals for different numbers of simulations~$N_{\rm sims}$ used to estimate the covariance matrix and for different posterior distributions: \hl, \sh\ or employing a frequentist-matching prior as introduced in \protect\cite{Percival2021} (abbreviated with ``PFSH'' in the table).}
\label{tab:credintervals}
\centering
\begin{tabular}{lccc}
\hline
$\mathbf{N_{\rm sims}}$ & \textbf{\hl} & \textbf{\sh} & \textbf{PFSH}\\
\hline
$400$  & 0.47 (0.84) & 0.58 (0.92) & 0.70 (0.95) \\
$600$  & 0.56 (0.89) & 0.63 (0.93) & 0.69 (0.96) \\
$1000$ & 0.60 (0.91) & 0.64 (0.94) & 0.69 (0.95) \\
$5000$ & 0.69 (0.94) & 0.69 (0.94) & 0.69 (0.95) \\
\hline
\end{tabular}
\end{table}

\section{Parameter Upper Limits}\label{sec:clscatter}

Both effects discussed previously, the scatter in MAP $r$~values (Sec.~\ref{sec:peakscatter}) and the underestimation of uncertainties (Sec.~\ref{sec:uncertain}), have a combined impact on the estimation of parameter \textit{upper limits}. The results in Sec.~\ref{sec:uncertain} suggest that, while the analytic marginalization assuming a Wishart distribution for the sample covariance matrix corrects for the likelihood from being randomly too wide or too narrow, there is still an increased scatter in the posterior peak positions, as seen in Sec.~\ref{sec:peakscatter}.\\ 

\begin{figure}
\centering
\includegraphics[width=\linewidth]{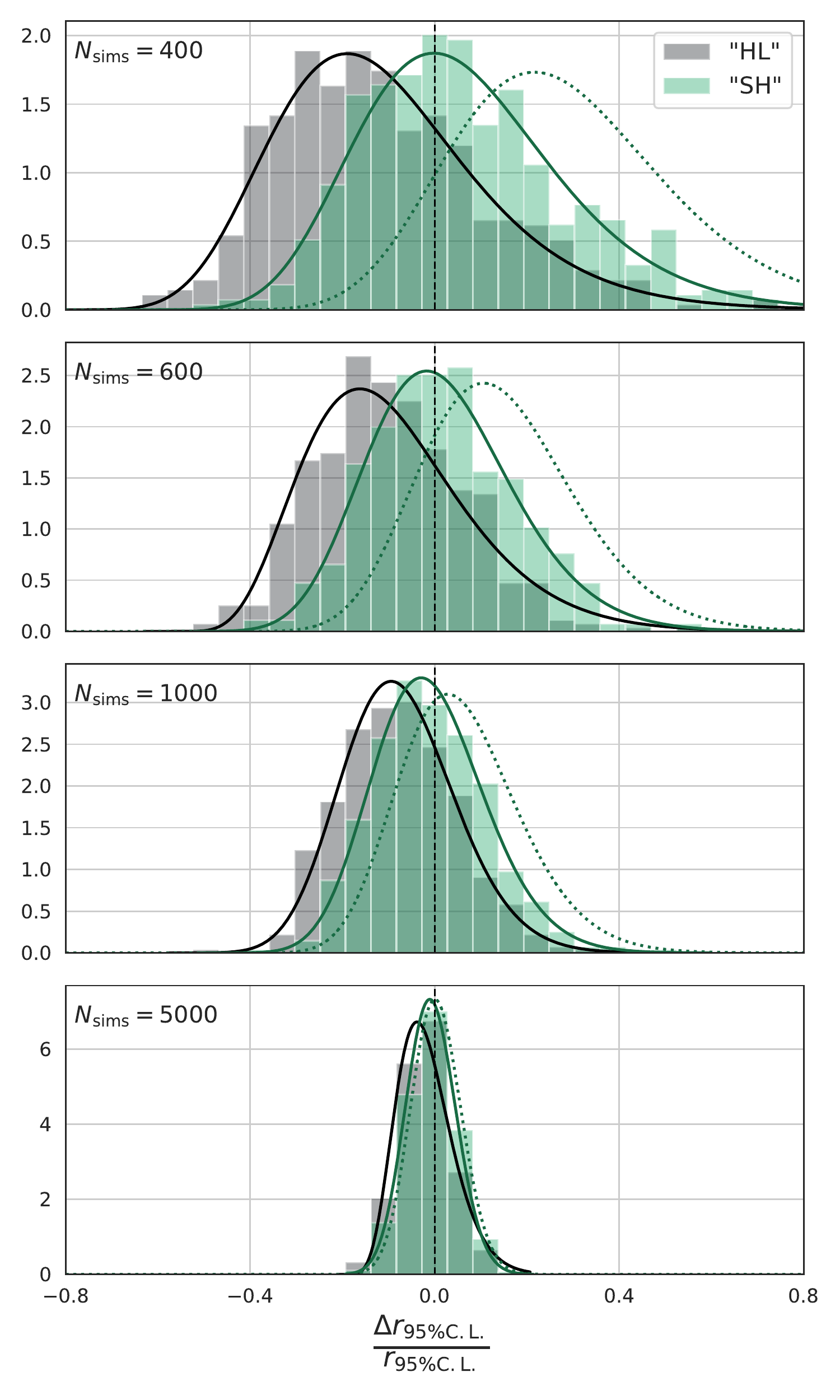}
\caption{The per-realization difference $\Delta r_{95\%\ \textrm{C.L.}}$ between the 95\%-C.L. upper limit for~$r$ estimated using a sample covariance matrix with $N_{\rm sims}=400,\ 600,\ 1000,\ 5000$ and using the fiducial ``true'' covariance; we normalize to the fiducial ``true'' upper limit. We find empirically that this difference is beta-prime distributed (in principle, resulting from a quotient of $\chi^2$-distributed variables), shown as a fit in the black solid line. We show the distribution for both likelihood cases, \hl\ and \sh. In the green dotted line, we show the variation of \sh\ employing a frequentist-matching prior \protect\citep{Percival2021} mentioned in Sec.~\ref{sec:uncertain}.} \label{fig:cl95}
\end{figure}

\begin{figure}
\centering
 \includegraphics[width=\linewidth]{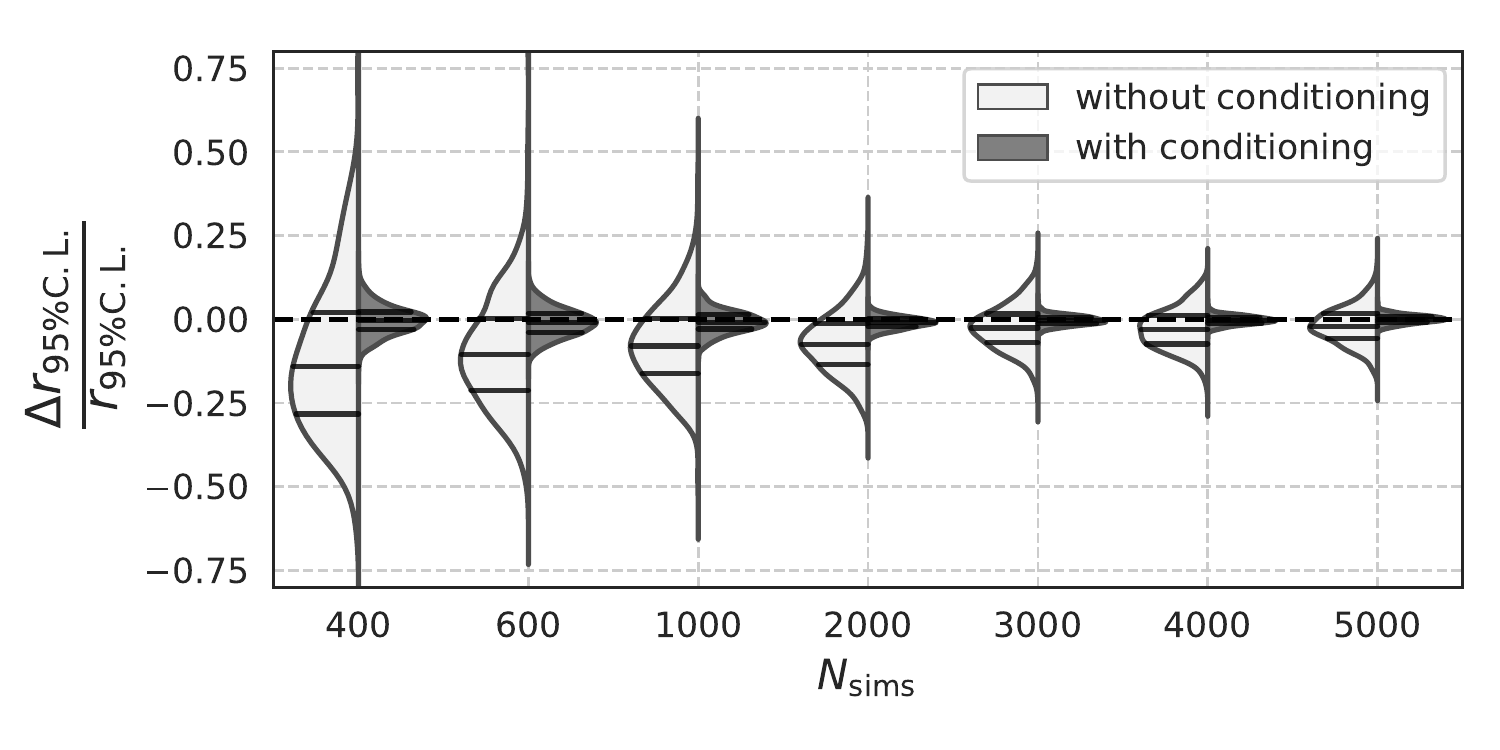}
\caption{The distributions of relative misestimations of the 95\% confidence limit on~$r$, as shown in Fig.~\ref{fig:cl95}, depending on~$N_{\rm sims}$. These are the kernel density estimates of the distributions for different~$N_{\rm sims}$ from the \hl\ likelihood along the respective vertical axis, either setting off-diagonal correlation matrix elements to zero (``with conditioning'') or not (``without conditioning''). The solid lines within the distributions indicate the quartiles of the respective distribution.} \label{fig:violin}
\end{figure}

In order to report an upper limit on~$r$, we restrict the prior to non-negative values. We compute a 95\%-C.L. upper limit for each of the simulated ``data'' realizations by imposing a uniform prior with $r\geq0$. In Figs.~\ref{fig:cl95} and \ref{fig:violin}, we show the distribution of relative differences between the estimated upper limit and the ``true'' upper limit, the former being obtained with a noisy covariance matrix with varying values of~$N_{\rm sims}$ and the latter with the ``true,'' effectively noiseless covariance matrix. For $N_{\rm sims}=400$, the upper limit is underestimated by more than~10\% in 58\%~of the realizations and by more than~30\% in 22\%~of the realizations. The \sh\ likelihood corrects for the uncertainty underestimation by accounting for the statistical uncertainty in the covariance matrix, but the scatter in the posterior peaks still causes an underestimation of the upper limit by more than~10\% in 24\%~of realizations. In addition, due to the shift to more positive values, the upper limit is misestimated by more than~10\% in 64\%~of realizations. 
We also include in Fig.~\ref{fig:cl95} the results from applying the SH likelihood variation with a frequentist-matching prior. In this case, because of the extended posterior widths, the upper limits are more often \emph{overestimated}, especially with small~$N_{\rm sims}$.
Moving our attention to the case of large~$N_{\rm sims}$, we find that the upper limits for $N_{\rm sims}=5000$ are underestimated by more than~10\% only 4\%~of the time for both likelihoods. We note that, for conditioned covariance matrices, a misestimation of the upper limit by more than~10\% happens only in less than~1\% of the cases for $N_{\rm sims}>400$, illustrated in Fig.~\ref{fig:violin} by the comparison of the two distributions for several representative values of~$N_{\rm sims}$.\\

\begin{figure}
\centering
 \includegraphics[width=\linewidth]{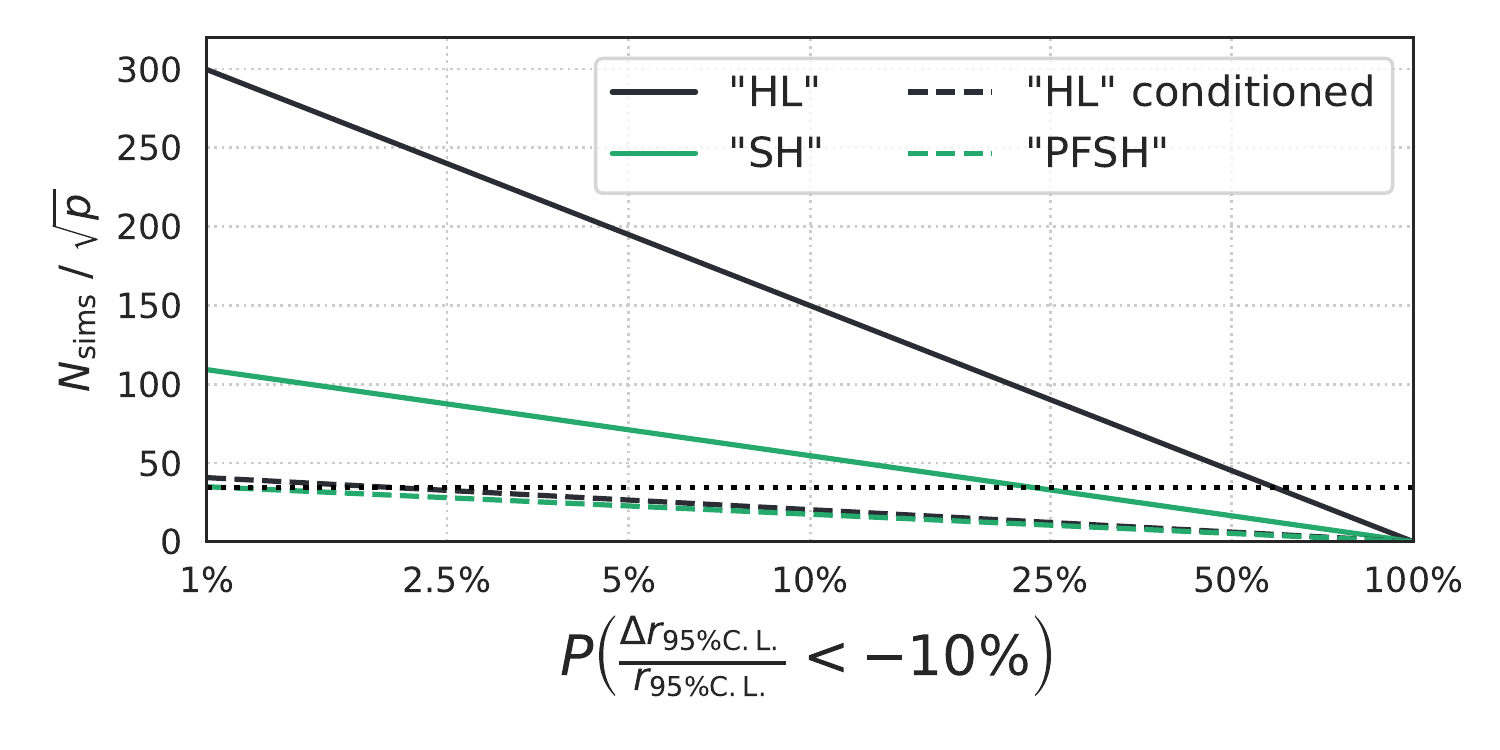}
\caption{Required number of simulations, $N_{\rm sims}$ per square root of the number of observables (i.e., bandpowers) $p$ to achieve a required precision on the estimate of the upper limit on $r$. This precision is given by the probability of underestimating the upper limit by more than 10\%. Different line colors indicate the considered likelihood approximation. The black dashed line corresponds to upper limits estimated with the \hl\ likelihood and a ``conditioned'' covariance matrix. In the case of the green dashed line we employ the frequentist-matching prior of \protect\cite{Percival2021} (``PFSH''). The dotted horizontal line indicates the baseline binning of this paper, $N_{\rm sims}=400$, which follows \protect\cite{Tristram20}. Due to being derived from a limited set of simulations, this figure should only be used for order-of-magnitude estimates.} \label{fig:nsimsp}
\end{figure}

In Fig.~\ref{fig:nsimsp} we provide a rule-of-thumb metric for choosing an appropriate number of simulations~$N_{\rm sims}$ for estimating an upper limit on~$r$ for an arbitrary number of bandpowers~$p$. To do so, we fit a beta-prime distribution as shown in Fig.~\ref{fig:cl95} to each empirically obtained distribution for various values of~$N_{\rm sims}$ and~$p$ as well as different likelihood approximations or conditioning schemes. For each fit, we evaluate the cumulative distribution function to obtain the probability that the upper limit on $r$ is underestimated by more than~10\%. We find empirically a logarithmic dependence between this probability and~$N_{\rm sims}/\sqrt{p}$. We fit for this logarithmic dependence and show it in Fig.~\ref{fig:nsimsp} for the \hl\ and \sh\ likelihood approximations as well as for the case of covariance-matrix conditioning and employing the frequentist-matching prior of \cite{Percival2021}. 
We find that in order to underestimate the upper limit on $r$ by more than 10\% in less than 5\% of cases, the ratio $N_{\rm sims}/\sqrt{p}$ is required to be larger than $195$ for the \hl\ likelihood and $71$ for the \sh\ likelihood. These numbers reduce to $26$ if conditioning the covariance matrix in the \hl\ likelihood or $23$ if employing the frequentist-matching prior in the \sh\ likelihood. The simulations included here use values of~$N_{\rm sims}$ between $200$ and~$5000$, $p$~between $39$ and~$135$, and the fixed cosmological and noise model described in Sec.~\ref{sec:toymodel}.

\section{Consequences for the \textit{Planck} ``\textit{lowlEB}'' likelihood}\label{sec:planck2020}

\begin{figure}
\centering
 \includegraphics[width=\linewidth]{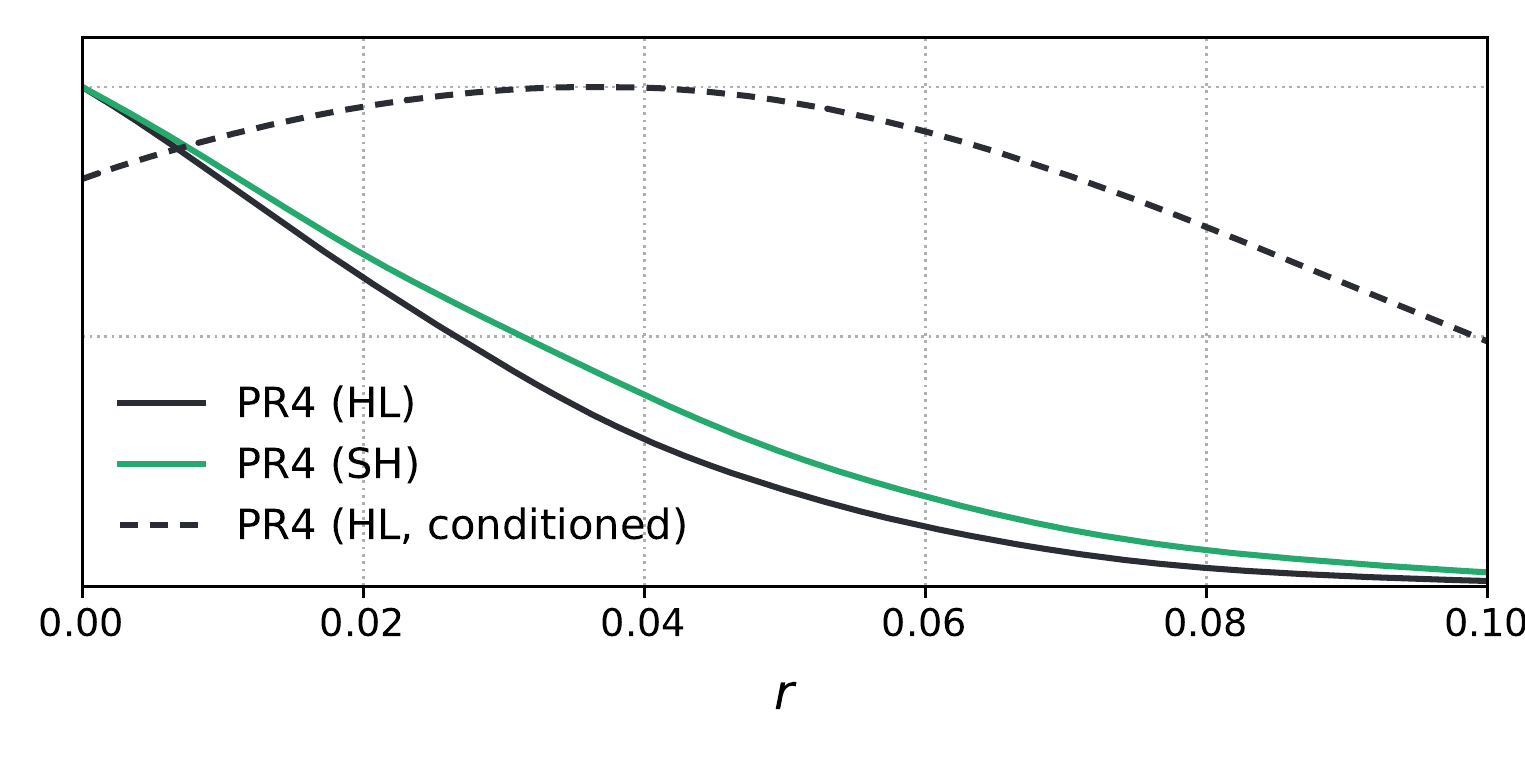}
 \includegraphics[width=\linewidth]{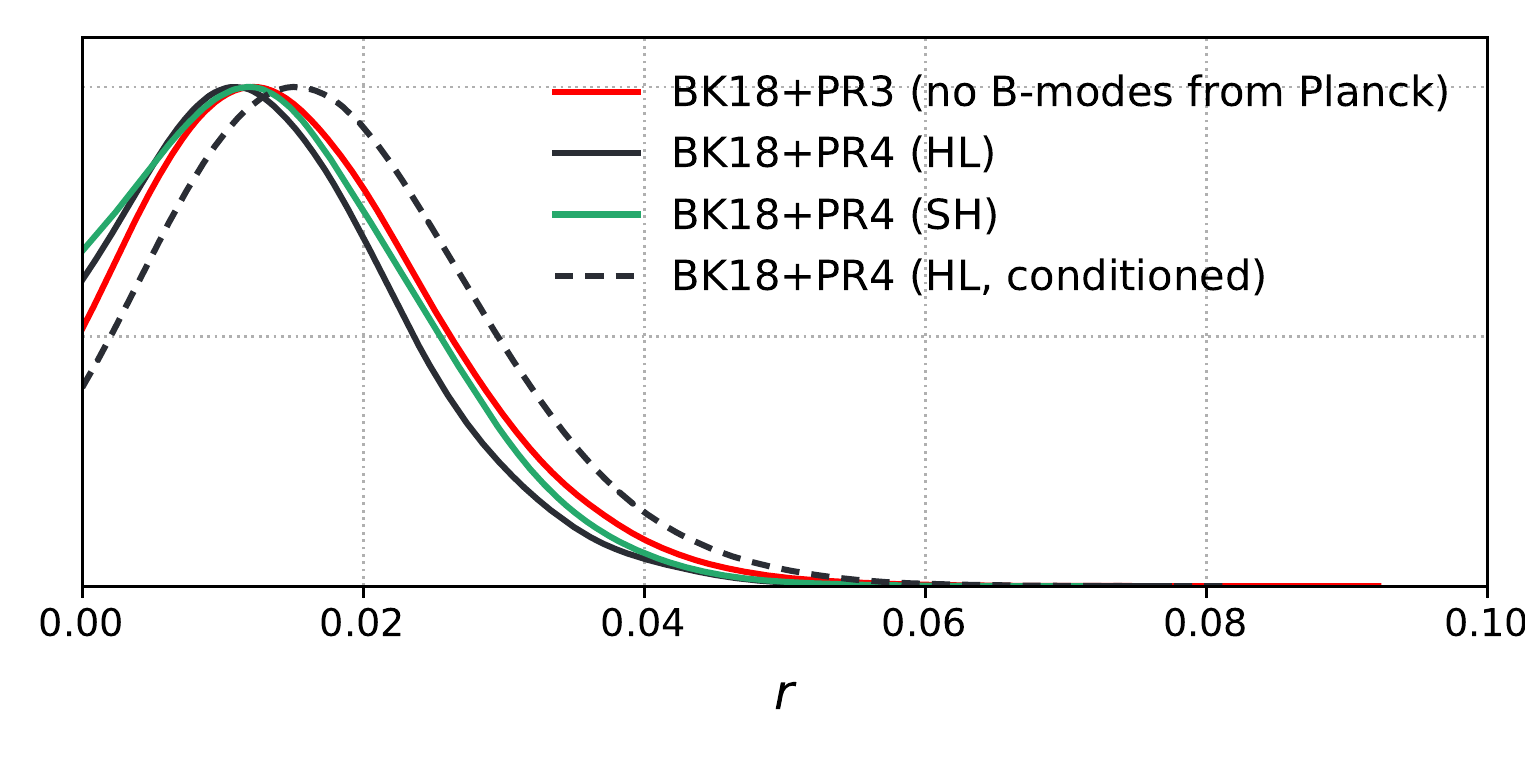}
\caption{Marginalized posterior densities for~$r$ for \textit{Planck} PR4 alone on the top and, on the bottom, for combinations of BK18 with \textit{Planck} data products: \textit{Planck} PR3 \citep{Planck18} as described in Fig.~5 of \protect\cite{BK18} and \textit{Planck} PR4 \citep{Planck20} as described in \protect\cite{Tristram21}. The latter combination is repeated for the different likelihood approximations discussed in Sec.~\ref{sec:parest} and with the conditioning scheme described in Sec.~\ref{sec:realdata} in the \textit{Planck} PR4 likelihood. The BK18 likelihood is used unchanged in all cases. The combined posteriors in the bottom plot include further information from \textit{Planck} 2018 lensing \citep{Plensing} and BAO \citep{bao} to be consistent with \protect\cite{BK18} and \protect\cite{Tristram21}.} \label{fig:realrposterior}
\end{figure}

To exemplify how MC~noise, as in the off-diagonal elements of Fig.~\ref{fig:toysimcov}, can cause shifts in posteriors estimated on real data, we attempt such conditioning in the \textit{Planck} ``\textit{lowlEB}'' likelihood. We deem the visible structure that we can observe in the matrix on the right-hand side of Fig.~\ref{fig:toysimcov} to be real. This includes bin-to-bin correlations over 2~bins for the low multipoles ($\ell\leq35$) and nearest-neighbor bin correlations for the higher multipoles. While we cannot rule out the possibility that there are real covariance structures beyond those matrix elements, circumstantial evidence suggests that their marginal distributions are consistent with MC~noise (cf. Sec.~\ref{sec:covmat}). As such, we apply conditioning to the \textit{Planck} ``\textit{lowlEB}'' covariance matrix by setting the ``MC-noise-dominated'' matrix elements to zero. This results in a large shift of the marginalized posterior of~$r$, shown in Fig.~\ref{fig:realrposterior}, and moves the 95\% C.L. upper limit from the ``\textit{lowlEB}'' likelihood alone reported in \cite{Tristram20} (``PR4 \hl'') to significantly larger values
\begin{subequations}\begin{align}
r_{0.05}&<0.069 \quad\mathrm{(PR4\ \hl)},\\
r_{0.05}&<0.075 \quad\mathrm{(PR4\ \sh)},\\
r_{0.05}&<0.13~~  \quad\mathrm{\mathbf{(PR4\ \hl,\ conditioned)}.}
\end{align}\end{subequations}
Combining the \textit{Planck} PR4 likelihoods with the recent results from BICEP/\textit{Keck} \citep[``BK18'']{BK18} we find the 95\% C.L. upper limits
\begin{subequations}\begin{align}
r_{0.05}&<0.035 \quad\mathrm{(BK18+PR3\ \hl)},\label{eq:bk18upperlimit}\\
r_{0.05}&<0.032 \quad\mathrm{(BK18+PR4\ \sh)},\label{eq:tristramupperlimit}\\
r_{0.05}&<0.038 \quad\mathrm{(BK18+PR4\ \hl,\ conditioned),\label{eq:trueupperlimit}}
\end{align}\end{subequations}
where the upper limit in Eq.~\ref{eq:bk18upperlimit} is taken from \cite{BK18} and obtained by combining BICEP/\textit{Keck} data with \textit{Planck} PR3 2018 high-$\ell$ TTTEEE, \textit{lowE} \citep{Planck18}, lensing \citep{Plensing} and BAO \citep{bao}. For the combinations of BICEP/\textit{Keck} with \textit{Planck} PR4 data we follow \cite{Tristram20} and include, apart from the ``lowlEB'' likelihood, \textit{Planck} PR4 high-$\ell$ TT of the \texttt{HiLLiPoP} likelihood\footnote{\texttt{https://github.com/planck-npipe/hillipop}} \citep{Couchot2017},  \textit{lowT} and lensing of \textit{Planck} PR3 2018 as well as BAO data. The upper limit in Eq.~\ref{eq:tristramupperlimit} is consistent with the one reported in \cite{Tristram21}.\\

We observe that the $r$ posterior of PR4 alone and in combination with BICEP/\textit{Keck} in the unconditioned case peaks at lower values compared to the case with covariance matrix conditioning. This suggests that the lower upper limits produced by the \textit{Planck} ``\textit{lowlEB}'' likelihood can be a result of the additional scatter observed in Secs.~\ref{sec:peakscatter} and~\ref{sec:clscatter}. More simulations ($\mathcal{O}(5\,000)$) are necessary to rule out a spurious misestimation for both used likelihood approximations, \hl\ and \sh. A similar conclusion can be drawn from an analysis of the same data set using a frequentist profile likelihood method \citep{Campeti2022}.

\section{Conclusions}\label{sec:realdata}

We address the issue of sample-covariance effects on the inference of the tensor-to-scalar ratio $r$ from CMB $E$- and $B$-mode bandpowers by performing parameter estimation on idealized simulated data. We employ likelihood-based methods that are standard in the field and used in several analyses of that type \citep{BK18,Tristram20,Tristram21,sptr,pbr}. \\

We observe an effect which is little addressed in the literature so far: the scatter of the posterior probabilities depending on the specific realization of the data bandpowers, which is beyond the expected fluctuation given the model. This introduces an additional scatter in the estimation of $r$ depending on the number of simulations used in the covariance-matrix estimation, which is not accounted for in standard parameter inference. In the case of our likelihood configurations with $\mathcal{O}(100)$ observables, we find that the effects can be sufficiently reduced with an increased number of simulations on the order of $\mathcal{O}(10\,000)$ or with appropriate matrix conditioning. \\

Monte Carlo (MC) noise in the estimated covariance matrix can cause a misestimation of parameter uncertainties, which can be effectively ameliorated by 
marginalizing over the probability distribution of the sample covariance matrix \citep{Sellentin15}. 
However, due to the aforementioned effect of realization-dependent statistical shifts of the posteriors, this does not prevent a misestimation of confidence limits nor of the best-fit maximum \textit{a~posteriori} value of~$r$.
This is particularly severe for the \hl\ likelihood paired with an unconditioned covariance matrix constructed with a limited number of simulations, which was implemented in \cite{Tristram20} with $N_{\rm sims}=400$, and is partially mitigated when the \sh\ likelihood is used as in \cite{Tristram21}, which also used $N_{\rm sims}=400$.
We conclude that, for a small number of simulations, an unconditioned covariance matrix can cause a non-negligible misestimation of a 95\%-C.L. upper limit. \\

Successful strategies in preventing this have employed some form of covariance-matrix conditioning \citep{Balkenhol21,Dutcher2021,BK18,sptr,pbr}. We show, however, that the published \textit{Planck} PR4 \texttt{LoLLiPoP} without a conditioned covariance matrix could be afflicted by the pathologies presented in this paper and resulting upper limits on $r$ are likely misestimated.  \\

Many experimental efforts are underway to improve our current constraints on the tensor-to-scalar ratio \citep{Moncelsi2020,SO,S4}. This work suggests that a careful treatment of the statistical inference method is necessary in order to conclude faithful and accurate statements from increasingly sensitive data sets. Similar considerations will have to be made for upper limits on other cosmological parameters such as the total mass of neutrinos \citep{Lesgourgues2012} and the fraction of early dark energy \citep{Poulin2019}.

\section*{Acknowledgements}
We thank Lennart Balkenhol, Neil Goeckner-Wald, Chao-Lin Kuo, Clem Pryke and Bryan Steinbach for many stimulating comments.
The results in this paper have been derived using \texttt{numpy} \citep{numpy}, \texttt{scipy} \citep{scipy}, \texttt{NaMaster} \citep{namaster}, the \texttt{cobaya} framework \citep{cobaya2021}, the \texttt{LoLLiPoP} \citep{mangilli2015,Tristram20} and \texttt{HiLLiPoP} \citep{Couchot2017} likelihoods as well as the \texttt{HEALPix}\footnote{\texttt{https://healpix.sourceforge.io/}} and \texttt{healpy} packages \citep{gorski2005,Zonca2019}. This research used resources of the National Energy Research Scientific Computing Center (NERSC), a U.S. Department of Energy Office of Science User Facility located at Lawrence Berkeley National Laboratory, operated under Contract No. DE-AC02-05CH11231 and the Odyssey cluster supported by the FAS Science Division Research Computing Group at Harvard University.
W.L.K.W is supported in part by the Department of Energy, Laboratory Directed Research and Development program and as part of the Panofsky Fellowship program at SLAC National Accelerator Laboratory, under contract DE-AC02-76SF00515.

\section*{Data Availability}
The data underlying this article will be shared on reasonable request to the corresponding author.



\bibliographystyle{mnras}
\bibliography{bib} 



\newpage
\appendix
\section{Definitions}\label{sec:appendix}
The function $g(x)$ introduced in Eq.~\ref{eq:defineX} is defined as
\begin{equation}
    g(x) \equiv {\rm sign}(x-1)\sqrt{2\left(x-{\rm ln}(x)-1\right)} .\label{eq:gx}
\end{equation}
Furthermore we employ the function ${\rm vecp}(\mathbf{A})$, which returns for a symmetric $n\times n$ matrix $\mathbf{A}$ the column vector
\begin{equation}
    {\rm vecp}(\mathbf{A})\equiv\left( A_{11},\ A_{21},\ ...,\ A_{n1},\ A_{12},\ A_{22},\ ...,\ A_{nn}  \right)^T. \label{eq:vecp}
\end{equation}
The normalization constant of the \sh\ likelihood is given by \citep{Sellentin15}
\begin{equation}
    c = \frac{\sqrt{2\pi}\Gamma\left(\frac{N}{2}\right)}{\left[\pi(N-1)\right]^{p/2}\Gamma\left(\frac{N-p}{2}\right)}, \label{eq:gammafct}
\end{equation}
where $N$ is the number of simulations used to extimate the covariance matrix, $p$ is the number of observables and $\Gamma$ is the Gamma function.


\bsp	
\label{lastpage}
\end{document}